\documentclass[epj]{svjour}
\usepackage{graphicx}
\def\sqeeb{\ifmmode{\sqrt{s_{\protect\bf\mathrm{ee}}}}\else
  {$\sqrt{s_{\protect\bf\mathrm{ee}}}$}\fi}
\def\epem{\ifmmode{\mathrm{e}^{+}\mathrm{e}^{-}}\else
  {$\mathrm{e}^{+}\mathrm{e}^{-}$}\fi}
\def\sqee{\ifmmode{\sqrt{s_\mathrm{ee}}}\else
  {$\sqrt{s_\mathrm{ee}}$}\fi}
\def\kp{{\ifmmode{k_{\perp}}\else{$k_{\perp}$}\fi}}
\def\etmean{\ifmmode{\bar{E}^{\mathrm{jet}}_{\mathrm{T}}}\else
  {$\bar{E}^{\mathrm{jet}}_{\mathrm{T}}$}\fi}
\def\etajmean{\ifmmode{|\bar{\eta}^\mathrm{jet}|}\else 
  {$|\bar{\eta}^\mathrm{jet}|$}\fi}
\def\lxg{\ifmmode{\mathrm{log_{10}}(x_{\gamma})}\else
  {${\mathrm{log_{10}}(x_{\gamma})}$}\fi}
\def\xg{\ifmmode{x_{\gamma}}\else{${x_{\gamma}}$}\fi}
\def\xgp{\ifmmode{x_{\gamma}^+}\else{${x_{\gamma}^+}$}\fi}
\def\xgm{\ifmmode{x_{\gamma}^-}\else{${x_{\gamma}^-}$}\fi}
\def\xgpm{\ifmmode{x_{\gamma}^{\pm}}\else 
  {${x_{\gamma}^{\pm}}$}\fi}
\def\ipb{\ifmmode {\mathrm{pb}^{-1}}\else 
  {$\mathrm{pb}^{-1}$}\fi}

\def\ee{\ifmmode{\mbox{e}^+\mbox{e}^-}\else
  {$\mbox{e}^+\mbox{e}^-$}\fi}
\def\thetamaxp{\ifmmode{\theta_\mathrm{max}'}\else
  {$\theta_\mathrm{max}'$}\fi}
\def\pt{\ifmmode{p_\mathrm{T}}\else{$p_\mathrm{T}$}\fi}
\def\Zzero{\ifmmode{\mathrm{Z}^{0}}\else{$\mathrm{Z}^{0}$}\fi}

\def\etjet{\ifmmode{E^\mathrm{jet}_\mathrm{T}}\else 
  {$E^\mathrm{jet}_\mathrm{T}$}\fi}
\def\ptmiss{\ifmmode{{P}_{\mathrm{T,MISS}}}\else 
  {${P}_{\mathrm{T,MISS}}$}\fi}
\def\mj1h2{\ifmmode{M_{\mathrm{J1H2}}}\else 
  {$M_{\mathrm{J1H2}}$}\fi}
\def\ebeam{\ifmmode{E_{\mathrm{BEAM}}}\else 
  {$E_{\mathrm{BEAM}}$}\fi}
\def\etajet{\ifmmode{|\eta^\mathrm{jet}|}\else 
  {$|\eta^\mathrm{jet}|$}\fi}
\def\etaj{\ifmmode{\eta^\mathrm{jet}}\else 
  {$\eta^\mathrm{jet}$}\fi}
\def\etajdef{\ifmmode{\eta^\mathrm{jet} = 
    -\ln\tan(\theta^\mathrm{jet}/2)}\else{$\eta^\mathrm{jet} = 
    -\ln\tan(\theta^\mathrm{jet}/2)$}\fi}
\def\detajet{\ifmmode{|\Delta\eta^\mathrm{jet}|}\else 
  {$|\Delta\eta^\mathrm{jet}|$}\fi}
\def\costhst{\ifmmode{|\mathrm{cos}\,\Theta^{*}|}\else 
  {$|\mathrm{cos}\,\Theta^{*}|$}\fi}
\def\etajetc{\ifmmode{|\eta^\mathrm{jet}_\mathrm{cntr}|}\else 
  {$|\eta^\mathrm{jet}_\mathrm{cntr}|$}\fi}
\def\etajetf{\ifmmode{|\eta^\mathrm{jet}_\mathrm{fwd}|}\else 
  {$|\eta^\mathrm{jet}_\mathrm{fwd}|$}\fi}
\def\etah{\ifmmode {\hat{\eta}}\else{$\hat{\eta}$}\fi}

\def\dsdeta{\ifmmode{\frac{\mathrm{d}{\sigma}_{\mathrm{dijet}}}
  {\mathrm{d}\etajet}}\else
    {$\frac{\mathrm{d}{\sigma}_{\mathrm{dijet}}}
      {\mathrm{d}\etajet}$}\fi}

\def\et{\ifmmode{E_\mathrm{T}}\else{$E_\mathrm{T}$}\fi}

\def\gg{\ifmmode{\gamma\gamma}\else{$\gamma\gamma$}\fi}
\def\gsg{\ifmmode{\gamma^{\star}\gamma}\else
  {$\gamma^{\star}\gamma$}\fi}
\def\PTMIA{\ifmmode{p_\mathrm{t}^\mathrm{mi}}\else
  {$p_\mathrm{t}^\mathrm{mi}$}\fi}
\def\sas1d{SaS\,1D}

\def\grv{GRV}
\def\grvnlo{GRV\,HO}

\def\gs96nlo{GS96\,HO}
\def\lac1{LAC\,1}
\def\hadcor{\ifmmode{(1+\delta_{hadr})}\else{$(1+\delta_{hadr})$}\fi}

\begin{document}
\title{Jet production in two-photon collisions at LEP}
\author{Thorsten Wengler\inst{1} % etc
}                     % Do not remove
\institute{CERN, EP-division, 1211 Geneva 23, Switzerland}
\date{Received: \hspace{1cm}/ Revised version: }
\abstract{
Jet and di-jet production are studied in collisions of quasi-real photons
collected during the LEP2 program at {\epem} center-of-mass energies
from 189 to 209\,GeV. OPAL reports good agreement of NLO
perturbative QCD with the measured differential di-jet cross
sections, which reach a mean transverse energy of the di-jet
system of 25\,GeV. L3, on the other hand, finds drastic disagreement of
the same calculation with single jet production for transverse jet
momenta larger than about 25\,GeV.
 % end of PACS codes
} %end of abstract
\maketitle
\section{Single jet inclusive production}

\begin{figure}
\label{sngl}
\includegraphics[width=0.45\textwidth]{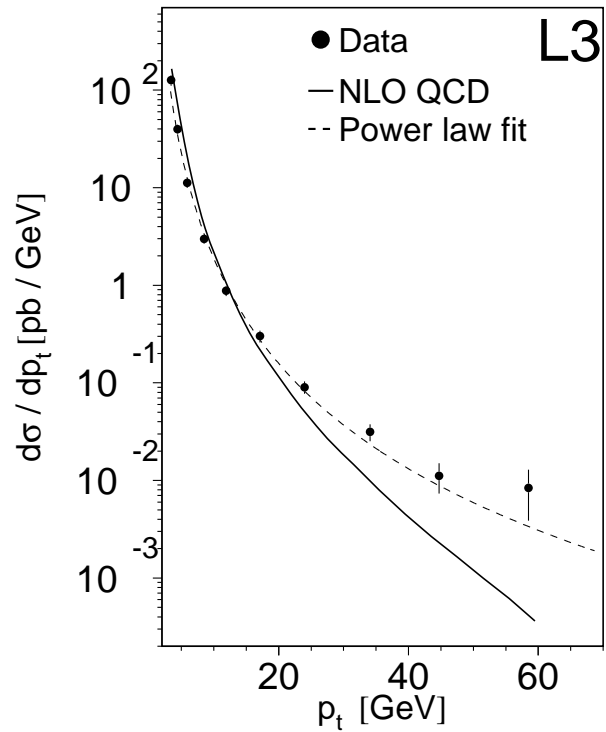}
\caption{Inclusive jet differential cross section as measured by L3
compared to NLO perturbative QCD calculations and the result of a
power law fit. The theoretical scale uncertainty is less than
20\%.}
\label{fig:1}      
\end{figure}

The L3 collaboration\,\cite{l3-det} has measured inclusive jet
production in photon-photon interactions\,\cite{l3-jets}. A total
integrated luminosity of 560\,{\ipb} recorded at {\epem}
center-of-mass energies $\sqee = 189-209$\,GeV is used, with a
luminosity weighted average of $\langle \sqee \rangle =
198$\,GeV. Photon-photon interactions in which one of the electrons
is scattered into the detector are rejected, such that both photons
are quasi-real. Jets are reconstructed using the {\kp}-clustering
algorithm\,{\cite{bib-ktclus}} and analysed in the pseudorapidity
range $|\eta| < 1$ for jet transverse momenta $3 < p_t <
70$\,GeV. The remaining background from other processes after event
selection increases from about 5\% at low $p_t$ to about 20\% at
high $p_t$. This background is subtracted bin-by-bin from
the data before corrections for selection efficiency and detector
acceptance are applied. The differential cross section as a
function of $p_t$ is shown in Figure\,\ref{fig:1}. The distribution
can be described by a power law function $Ap_{t}^{-B}$ with
$B=3.6\pm0.1$. A comparison to a NLO perturbative QCD
calculation\,\cite{bib-frixber} using the {\grvnlo} parton density
functions\,\cite{bib-grv} and $\Lambda^{\left(5\right)}=152$\,MeV
predicts a much softer spectrum, as can be seen in
Figure\,\ref{fig:1} and fails to describe the data for jet
transverse momenta larger than about 25\,GeV.

\section{Di-jet production and jet structure}

OPAL\,\cite{det-opal} has studied the production of di-jets in the
collisions of two quasi-real photons at an {\epem} centre-of-mass
energy {\sqee} from 189 to 209\,GeV, with a total integrated
luminosity of \mbox{593\,\ipb}. Di-jet events are of particular
interest, as the two jets can be used to estimate the fraction of
the photon momentum participating in the hard interaction, {\xg},
which is a sensitive probe of the structure of the photon. Also
here the {\kp}-clustering algorithm\,{\cite{bib-ktclus}} is used for
the measurement of the differential cross-sections, because of the
advantages of this algorithm in comparing to theoretical
calculations\,{\cite{bib-ktisbest}}. The cone jet algorithm is used
to study the different structure of the cone jets compared to
jets defined by the {\kp}-clustering algorithm.

In LO QCD, neglecting multiple parton interactions, two hard parton
jets are produced in {\gg} interactions.  In single- or
double-resolved interactions, these jets are expected
to be accompanied by one or two remnant jets.  A pair of variables,
{\xgp} and {\xgm}, can be defined  that estimate
the fraction of the photon's momentum participating in the hard
scattering:
\begin{equation}
\xgpm \equiv \frac{\displaystyle{\sum_{\rm jets=1,2}
 (E^\mathrm{jet}{\pm}p_z^\mathrm{jet})}}
 {{\displaystyle\sum_{\rm hfs}(E{\pm}p_z)}} ,
\label{eq-xgpm}
\end{equation}
where $p_z$ is the momentum component along the $z$ axis of the
detector and $E$ is the energy of the jets or objects of the
hadronic final state (hfs).  In LO, for direct events, all energy
of the event is contained in two jets, i.e.,~${\xgp}=1$ and
${\xgm}=1$, whereas for single-resolved or double-resolved events
one or both values are smaller than~1. Differential cross sections
as a function of {\xg} or in regions of {\xg} are therefore a
sensitive probe of the structure of the photon.

\subsection{Jet structure}
The internal structure of jets is studied using the jet
shape, $\Psi(r)$, which is defined as the fractional transverse jet
energy contained in a subcone of radius $r$ in $\eta$-$\phi$ space
concentric with the jet axis, averaged over all jets of the event
sample.  Both {\kp} and cone jets are analysed in this way. As
proposed in \cite{bib-seym}, only particles assigned to the jet by
the jet finders are considered. Events entering the jet shape
distributions are required to have at least two jets with a
transverse energy 3\,GeV $< {\etjet} <$ 20\,GeV and a pseudo-rapidity
${\etajet} < 2$.
\begin{figure}[htb]
\includegraphics[width=0.47\textwidth]{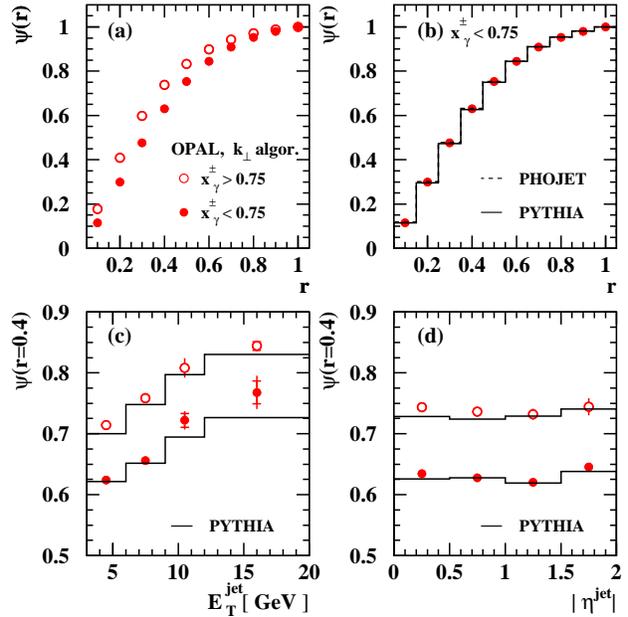}
\caption{The jet shape, $\Psi(r)$, for the two regions of 
  {\xgp}-{\xgm}-space indicated in the figure (a), and $\Psi(r)$ for
  $\xgpm < 0.75$ compared to the predictions of the LO MC generators
  PHOJET and PYTHIA (b). Figures (c) and (d) show the value of
  $\Psi(r=0.4)$ as a function of the transverse energy and
  pseudo-rapidity of the jet respectively, compared to the PYTHIA
  prediction. }
\label{fig:jsh01}
\end{figure}
\begin{figure}[htb]
\includegraphics[width=0.47\textwidth]{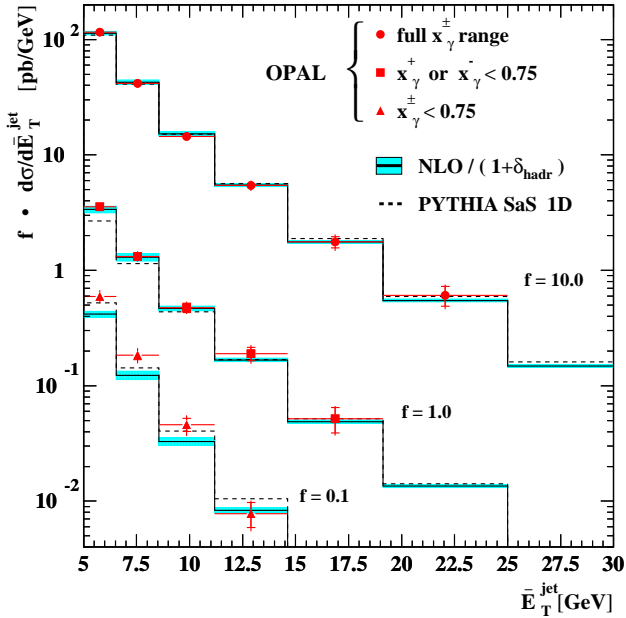}
\caption{The di-jet cross-section as a function of the mean
  transverse energy $\etmean$ of the di-jet system, for the three
  regions in {\xgp}-{\xgm}-space given in the figure.
  The factor $f$ is used to separate the three measurements in the
  figure more clearly.}
\label{fig:etmxs}
\end{figure}

In Figure\,\ref{fig:jsh01}\,(a) the jet shape, $\Psi(r)$, is shown
for the {\kp} algorithm for both ${\xgpm} > 0.75$ and ${\xgpm} <
0.75$. The first sample is dominated by direct photon-photon
interactions and hence by quark-initiated jets. As is demonstrated
in the figure, jets in this sample are more collimated than for
small values of {\xgpm}, where the cross-section is dominated by
resolved processes and hence has a large contribution from
gluon-initiated jets. In both cases the jets become more collimated
with increasing transverse energy, as is shown in
Figure\,\ref{fig:jsh01}\,(c). There is no significant dependence on
the jet pseudo-rapidity (Figure\,\ref{fig:jsh01}\,(d)).  Both
PHOJET\,\cite{bib-phojet} and PYTHIA\,\cite{bib-pythia} give an
adequate description of the jet shapes as can be seen in
Figures\,\ref{fig:jsh01}\,(b), (c), and (d). The default choices of
{\sas1d}\,\cite{bib-sas} for PYTHIA and LO {\grv}\,\cite{bib-grv} for
PHOJET are taken. Comparisons of jets defined by the cone
algorithm and the {\kp} algorithm (not shown here) lead to the
conclusion that the behavior described above is similar for both
jet algorithms, however cone-jets are significantly broader than the jets
defined by the {\kp} algorithm at low {\etjet}. With increasing
{\etjet}, jets become more collimated and the two jet algorithms
become similar.

\subsection{Differential Di-jet cross-sections}
Only the {\kp} jet algorithm is used for the measurement of the
differential di-jet cross-sections. The experimental results are
compared to a perturbative QCD calculation at NLO\,\cite{bib-ggnlo}
which uses the {\grvnlo} parametrisation of the parton distribution
functions of the photon\,\cite{bib-grv}. The renormalisation and
factorisation scales are set to the maximum $\etjet$ in the event.
The calculation was performed in the
$\overline{\mathrm{MS}}$-scheme with five light flavours and
$\Lambda^{(5)}_{\mathrm{QCD}}=130$\,MeV. This calculation was shown
to be in agreement with the calculation compared to the single
inclusive jet measurement above for the di-jet observables
presented here\,\cite{bib-frixber}.  The average of the
hadronisation corrections estimated by PYTHIA and HERWIG have been
applied to the calculation for this comparison.  In the figures
described below the shaded band indicates the theoretical
uncertainty estimated by the quadratic sum of two contributions:
variation of the renormalisation scale by factors of 0.5 and 2 and
the difference between using HERWIG or PYTHIA in estimating the
hadronisation corrections.

The differential di-jet cross-section as a function of the mean
transverse energy $\etmean$ of the di-jet system is shown in
Figure\,{\ref{fig:etmxs}}.  At high
{\etmean} the cross-section is expected to be dominated by direct
processes, associated with the region ${\xgpm} > 0.75$.  Consequently
we observe a significantly softer spectrum for the case ${\xgpm} <
0.75$ than for the full {\xgp}-{\xgm}-space. The calculation is in
good agreement with the data for the full {\xgp}-{\xgm}-range and for
{\xgp} or {\xgm} $< 0.75$. The cross-section predicted for ${\xgpm} <
0.75$ is below the measurement. It should be noted that in this region the
contribution from the underlying event, not included in the
calculation, is expected to be largest, as shown
below. PYTHIA 6.161 is in good
agreement with the measured distributions using the {\sas1d} parton
densities. PYTHIA includes a model of the underlying event using
multiple parton interactions (MIA). 
\begin{figure}[htb]
\includegraphics[width=0.47\textwidth]{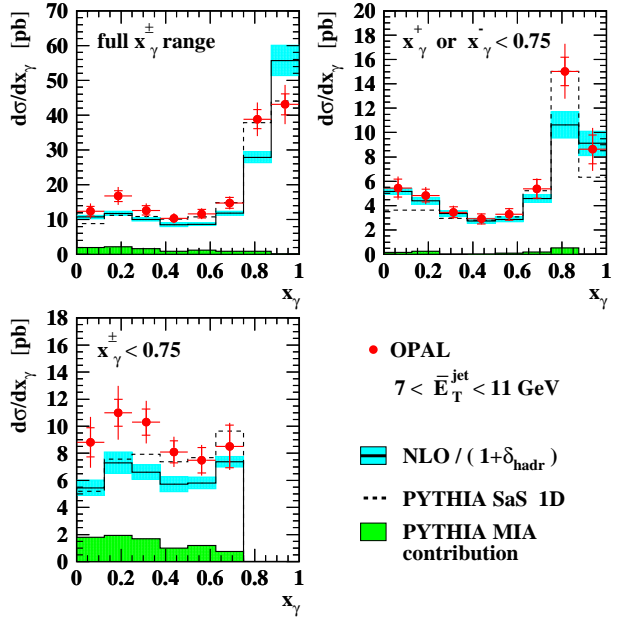}
\caption{The di-jet cross-section as a function of $\xg$ and for
the regions of the mean transverse energy $\etmean$ and {\xgpm} of
the di-jet system indicated in the figures. }
\label{fig:xgcomb}
\end{figure}

The three plots of Figure\,{\ref{fig:xgcomb}} show the differential
cross section as a function of {\xg} for the three regions in
{\xgp}-{\xgm}-space described above. The shaded histogram on the
bottom of each of the three plots indicates the contribution of MIA
to the cross section as obtained from the
PYTHIA\,{\cite{bib-pythia}} MC generator. It is evident especially
for ${\xgpm}<0.75$ that the MIA contribution is of about the same size
as the discrepancy between the measurement and the NLO
prediction. Furthermore it is interesting to observe that there is
next to no MIA contribution to the cross section if either {\xgp}
or {\xgm} is required to be less than one, while the sensitivity to
the photon structure at small {\xg} is retained. As one would
expect also the agreement of the NLO calculation with the
measurement is best in this case. For large {\xg} the NLO
calculation does not agree well with the data. However, it has been
pointed out that the calculation of the cross section becomes
increasingly problematic when approaching
${\xg=1}$\,\cite{bib-frixber,bib-klasenrev}.

With these measurements one is
therefore able to disentangle the hard subprocess from soft
contributions and make the firm statement that NLO perturbative QCD
is adequate to describe di-jet production in photon-photon
collisions in the regions of phase space where the calculation can
be expected to be complete and reliable, i.e. where MIA
contributions are small and for {\xg} not too close to unity. At
the same time a different sub-set of observables can be used to
study in more detail the nature of the soft processes leading to
the underlying event.


\end{document}